 \documentclass[prl,aps,twocolumn,showpacs]{revtex4}
\usepackage[dvips]{graphicx}
\usepackage{dcolumn}
\newcommand{\be}{\begin{equation}}
\newcommand{\ee}{\end{equation}}
\newcommand{\bea}{\begin{eqnarray}}
\newcommand{\eea}{\end{eqnarray}}
\newcommand{\nn}{ \nonumber}

\begin{document}

\title{Microwave Rectificatin of the Current at the Metal-Metal Junction
for Dilute 2D Metals}

\author{\bf Natalya A. Zimbovskaya}

\affiliation{Department of Physics, The City College of CUNY, New York, NY,
10021}


 \begin{abstract} 
Within a quasiclassical transport theory for 2D electron system we analyze
a recently observed effect of microwave rectification at the boundary
between two 2D metals of different carrier densities. Nonlinear response
is employed to explain the effect. It is shown that the
effect of rectification arises due to inhomogeneity of the electron
density which breaks space symmetry of the system. The results agree with
the above experiments.

\end{abstract}


\maketitle

 At present there is an increasing interest in studies of nonlinear
effects in transport phenomena. These studies may lead to many novel
results and have important applications such as the unidirectional
transport of molecular motors in the biological realm [1] and other kinds
of noise induced chaotic transport [2], and electronic transport through
superlattices [3], to mention a few. It is shown that when some space-time
symmetries characterizing a system under study are broken, this can give
rise to a nonlinear response of such system to an external disturbance
[1,4-8].

 The present work is motivated with the results of recently reported
experiments [9]. A nonlinear response of a dilute 2D electron system of a
significantly nonuniform electron density $n \bf (r)$ to an alternating
electric field $ {\bf E}(t)$ including the rectification of the current
was observed in these experiments. The inhomogeneity in the electron
density was produced in the experiments by the applied gates voltage. As a
result the density was monotonically varying along a direction
perpendicular to the lines of the applied electric potential taking on
values between $ \sim 2 \times 10^{11}$ and $ \sim 8 \times 10^{11}
$cm$^{-2}.$ Here we show that the effect of rectification can be
provided by the  electron density gradient $ \nabla n \bf (r)$.

 It is well known that when two conductors of different carrier densities,
and, therefore, of different work functions are put in contact, the
contact has a nonlinear current voltage characteristic which provides the
rectification of high-frequency currents flowing through the
interface. The theory
of the rectification properties of metal-metal junctions is well developed
(See e.g. [10]). Usually the rectification is attributed to electron
tunneling through the potential barrier at the metal-metal
interface. However, semiclassical analysis of the rectification through a
metal-metal interface, based on the Boltzmann transport equation also was
carried out [11--13]. These theories are applicable when a width of the
junction is small compared to characteristics of the electron transport.

 However, in the experiments of [9] both mean free path of electrons $ l $
and the width $ d $ of the slit separating the regions of smaller electron
density $n_1 $ and larger density $n_2 $ are of the same order
$(10^{-5}$cm). More thorough estimation showes that in these experiments $ 
d \geq l, $ and the ratio $ d/l $ could take on values close to
5. Therefore we consider the case when the electron density varies slowly
enough, so that the length of the interval $ d $ where $ \nabla n $
significantly differs from zero is large compared to the mean free path of
electrons. As a first approximation we assume that $ d >> l $ in further
analysis.

 To simplify following calculations we introduce a simple model where a
pair of 2D metals contiguous through the conducting boundary is replaced
by a single metal of strongly inhomogeneous electron density.  Then both
electron density and the chemical potential $ \mu \bf (r)$ depend on {\bf
''r''} $ (\mu ({\bf r}) = \displaystyle{{n(\bf r)}/{D_0}},$ where $ D_0 $
is the electron density of states at the Fermi surface).

 We treat electrons included into the considered 2D system as
noninteracting quasiparticles obeying classical dynamics and
subject to external forces and friction. Their motion is
described with the equation
 \be
\frac{d \bf p}{d t} + e \nabla V ({\bf r}) - e {\bf E } (t) + \gamma {\bf
p} = 0.
           \ee 
 Here, $ \bf p $ is the electron quasimomentum, and $ V \bf (r)$ is the
applied voltage producing the inhomogeneity in the electron density. The
friction term in the equation of motion (1) is written in a simplest form
$ \gamma \bf p $ where $ \gamma $ is the characteristic relaxation
frequency of the electron system. In fact, this term describes changes in
the quasimomentum of a single electron due to its scattering.  When there
is no alternating electric field $ {\bf E} (t) $ the electron system is in
the local equilibrium maintained with an internal electrochemical field $
\displaystyle{\frac{1}{e} \nabla \mu ({\bf r}) = {\nabla n (\bf r)}/{e
D_0} }$ which balances the external field $ - \nabla V \bf (r).$ The
presence of the field $ {\bf E}(t) $ violates this balance. However, when
the alternating contribution to the external electric field $ {\bf E}(t) $
is weak compared to the static contribution $ - \nabla V \bf (r) ,$ the
latter is still nearly equal to the electrochemical field arising in the
2D electron system, and we can replace the second term in the equation (1)
by $ \displaystyle{{\nabla n\bf (r)}/{D_0}}.$

Assuming for simplicity that $ {\bf E} (t) $ is directed in parallel with
the electron density gradient (along the "x" axis in a chosen coordinate
system) we rewrite the equation (1) in the form
 \be
 \frac{d^2 x}{d t^2} + \gamma \frac{dx}{dt} - \frac{1}{mD_0} \frac{dn}{dx}
- \frac{e}{m} E(t) = 0
           \ee 
 where $m$ is the electron effective mass. Symmetries of such equations
were analyzed in detail in earlier works [5-8]. It was shown there that
for asymmetric space dependent force $\big( $ in our case this force
equals $\displaystyle{\frac{1}{D_0} \frac{d n}{d x}} \big )$ system of
quasiparticles obeying equation of motion of the form (2) could give a
nonlinear and asymmetric adiabatic response to the ac field $ E (t) $
including a dc contribution to the current regardless of symmetries of
$E(t) $ if dissipation is included $(\gamma \neq 0). $ Whithin the
dissipationless limit $ (\gamma = 0) $ the effect of rectification
disappears when the ac field possesses $E_s $ symmetry defined as $ E (t +
\theta) = E (- t + \theta) $ even in the case when the spatial symmetry is
broken. However, in further analysis we rule this case out assuming that
the dissipation constant $ \gamma $ takes on nonzero values.

In the experiments of [9] the applied gates voltage produced an aperiodic
profile of the electron density as it depends on space
coordinates. Therefore the density gradient $ \nabla n \bf (r) $ also is
aperiodic. Assuming that $ \nabla n \bf (r) $ is asymmetric (and there
are no grounds for otherwise anticipations) we can expect the
rectification of the current to be revealed in the response to the
microwave field $ {\bf E}(t) $ for $ \gamma \neq 0.$

To get the desired response of the electron system we use the Boltzmann
transport equation for the electron distribution function $f( {\bf r,
p},t) $ taken within the relaxation time approximation:
 \be
\frac{\partial f}{\partial t} +  \frac{{\bf v}\partial f}{\partial \bf r}
+
\frac{d \bf p}{d t} \frac{\partial f}{\partial \bf p} =
- \frac{f - f_{eq} }{\tau}
       \ee
where the function $ f_{eq}$ describes the local equilibrium state of the
electron system in the absence of the microwave radiation $ E(t)$ and
equals the Fermi distribution function:
  \be
f_{eq} = f_0 (\varepsilon, \mu ({\bf r}), T).
     \ee
 Here, $ \epsilon = p^2/2m $ is the quasiparticle energy. In further
calculations it is supposed that the relaxation time $ \tau $ is finite
and nonzero to provide dissipation to be included [14].

Following [4] we expand $ f( {\bf r, p}, t)$ in harmonic polynomials, 
keeping three first terms of the expansion:
 \be
f({\bf r,p}, t) = f_{eq} + f_1^\alpha p_\alpha + f_2^{\alpha\beta}
\bigg(p_\alpha p_\beta - \frac{p^2}{2} \delta_{\alpha \beta}\bigg). 
    \ee
 The coefficients $f_1^\alpha, f_2^{\alpha\beta} $ depend on the
difference $ \varepsilon - \mu (\bf r) $, and the time; $ \alpha, \beta =
x,y.$

 Such approximation is justified for a weak microwave field because next
terms of the expansion give smaller corrections in terms of the microwave
field magnitude. Substituting this expansion into Boltzmann equation we
get a set of equations:
     $$
\frac{\partial f_1^\alpha}{\partial t} +
\frac{e}{m}E_\alpha(t) \frac{\partial f_{eq}}{\partial \varepsilon} +
2e \bigg (E_\beta (t) + \frac{1}{e} \frac{\partial \mu}{\partial r_\beta}
 \bigg ) f_2^{\beta \alpha}
      $$
   \be
- e \frac{p^2}{2m} E_\alpha (t)
\frac{\partial f_2^{\alpha \beta}}{\partial \varepsilon}
\delta_{\alpha \beta} = - \frac{f_1^\alpha}{\tau};
           \ee
          \be
\frac{\partial f_2^{\alpha\beta}}{\partial t} +
\frac{e}{m}E_\alpha(t) \frac{\partial f_{1}^\beta}{\partial t} =     
- \frac{f_2^{\alpha \beta}}{\tau}.
      \ee
 Here the term $\displaystyle{\frac{1}{e} \frac{\partial \mu}{\partial
r_\beta} = \frac{1 }{e D_0} \frac{\partial n}{\partial r_\beta} }$
describes the internal ''electrochemical'' field
arising due to inhomogeneity of the electron density. 
 To proceed we take the microwave field $ {\bf E}(t)$ of frequency $
\omega $ as:
     \be
 {\bf E} (t) = \frac{1}{2} ({\bf E}_\omega e^{i \omega t} + 
{\bf E}_\omega^* e^{-i \omega t})
        \ee
 Here, $ {\bf E}_{\omega} = {\bf E}'_{\omega} + i {\bf E}''_{\omega} $.
Further we assume
that $ \omega \tau < 1$ to provide the adiabatic response. This does not
contradict the relevant experiments of [9]. Symmetry propeties of the
Boltzmann transport equation were analyzed before [7], and it was shown
that they completely agree with the symmetries of the initial equations of
motion of relevant particles. In the present analysis we stipulate that
the density gradient $ \nabla n \bf (r) $ is asymmetric and aperiodic and
the relaxation time is finite, so we can expect the current rectification
to be exhibited.


We expand the coefficients $ f_1^\alpha, f_2^{\alpha,\beta} $ in time
Fourier series. Then, solving the system (6), (7) by perturbation theory
we present the current density $\bf j:$
      \be
j_\alpha = \frac{2 e}{m^2 (2\pi \hbar)^2} \int f_1^\beta p_\beta p_\alpha
d^2 p 
  \ee
 as an expansion in powers of the microwave magnitude $ E_\omega.$ Besides
the term which describes a linear response of the electron system to the
alternating field, this expansion includes a dc term $ {\bf j}_{(1)}$
corresponding to the rectified current density: 
 \be
{\bf j}_{(1)} = - \frac{e^3 \tau^3}{m^2} \, 
\frac{{\bf R} + \omega \tau
\bf Q}{1 + (\omega \tau)^2}
           \ee 
 and the term ${\bf j}_{(2)} $ which describes the current density at the
doubled frequency $ 2 \omega $:
 \bea
{\bf j}_{(2)} = - \frac{e^3 \tau^3}{m^2} \,
\frac{1}{(1 + (\omega \tau)^2)(1 + 4 (\omega \tau )^2 )^2}
 \nn \\ \nn \\
\times \big \{\big(1 - 8 (\omega \tau)^2 \big) 
\big({\bf R} \cos 2 \omega t - {\bf Q} \sin 2 \omega t\big)
 \nn \\ \nn \\
 + 
\omega \tau \big (5 - 4(\omega \tau)^2 \big )
\big({\bf R} \sin 2 \omega t + {\bf Q} \cos 2 \omega t\big) \big \}
  \eea
 where

  \bea
 {\bf R} ({\bf E}_{\omega}) =
{\bf E}'_{ \omega} ({\bf E}'_{ \omega} \cdot \nabla n )
+ {\bf E}''_{ \omega} ({\bf E}''_{ \omega} \cdot \nabla n )
 \nn \\ \nn \\
 {\bf Q} ({\bf E}_{\omega})=
{\bf E}''_{ \omega} ({\bf E}'_{ \omega} \cdot \nabla n )
- {\bf E}'_{ \omega} ({\bf E}''_{ \omega} \cdot \nabla n ) .
 \eea
 Both nonlinear contributions are quadratic in $ { E}_\omega ,$ and they
originate from the inhomogeneity in the electron density. When $ \nabla n$
goes to zero, the rectification disappears. Within a low frequency limit $
\omega \tau << 1$ the nonlinear contributions to the currents (10), (11)
do not vanish. This brings an asymmetry into the current-voltage
characteristics of the unhomogeneous electron system, even for $ \omega
\to 0. $ Similar results were obtained in earlier works for different
mechanisms of the rectification.
 We also can estimate an average of the rectified contribution to the
current density over the space interval $ d.$ The averaged quantities $
<{\bf R (E}_{\omega})>$ and $<{\bf Q (E}_{\omega})>$ are proportional to $
(n_2 - n_1)/d.$ This confirms that an asymmetric profile of the electron
density can give rise to the effect of the current rectification. However,
the obtained results are valid only for small magnitudes of the microwave
field, when the quadratic in $ { E}_\omega $ terms in the expansion for
the current density are smaller than the linear term.Therefore the
asymmetry of the current-voltage characteristic at the low-frequency limit
may be unavailable for observations.

 The nonlinear contributions to the current densities ${\bf j}_{(1)}, {\bf
j}_{(2)} $ occur when the microwave field has a component directed
perpendicularly to the boundary between 2D metals. Otherwise they both
turn zero. It also follows from (10)-(12) that the direction of the
currents reverses when we reverse the polarity of the gates, and,
consequently $ \nabla n.$ This agrees with the experimental results [9].

To make a comparison of the present results with the experiments [9] we
roughly estimate a voltage $ V_{dc} ,$ which corresponds to the rectified
current. Assuming that the microwave field is applied along the ''x''
direction as well as $ \nabla n $ we arrive at the estimation:
   \be 
 |V_{dc}| = \left | \int_{- \infty}^\infty \frac{j_{dc}(x)}{\sigma_0(x)}dx
\right |.
          \ee
 Here $\sigma_0 (x)$ is the Drude conductivity  for the electron
density $ n(x) $ and relaxation time $ \tau (x).$

It was reported [9] that when the minimum $ n(x)$ is larger than a certain
value $ n_0 \; (n_0\approx 2.5 \times 10^{11} \frac{1}{cm^2})$ the
relaxation time $\tau $ is nearly independent on the electron
concentration and can be treated as a constant in carrying out integration
over ''x''in (11). Using this approximation we get:
           \be
|V_{dc}| \approx \frac{4}{\pi} \frac{e}{m} \tau^2 \ln
\frac{n_1}{n_2}E_\omega^2.
      \ee
 The above estimate provides a reasonable agreement with the results of
[9] concerning the dependence of the rectification signal on the electron
densities of contiguous 2D-metals, as shown in Fig.1. Plotting curves in
this figure we assumed in accordance with [9] that $ m \approx 0.2 m_0 \;
(m_0 $ is the mass of a free electron); $ \tau \approx 3 \times 10^{-12}
s^{-1}$ and $ E_\omega \approx 10^3 \frac{V}{m}.$

\begin{figure}[t]
\begin{center}
\includegraphics[width=7.0cm,height=5.7cm]{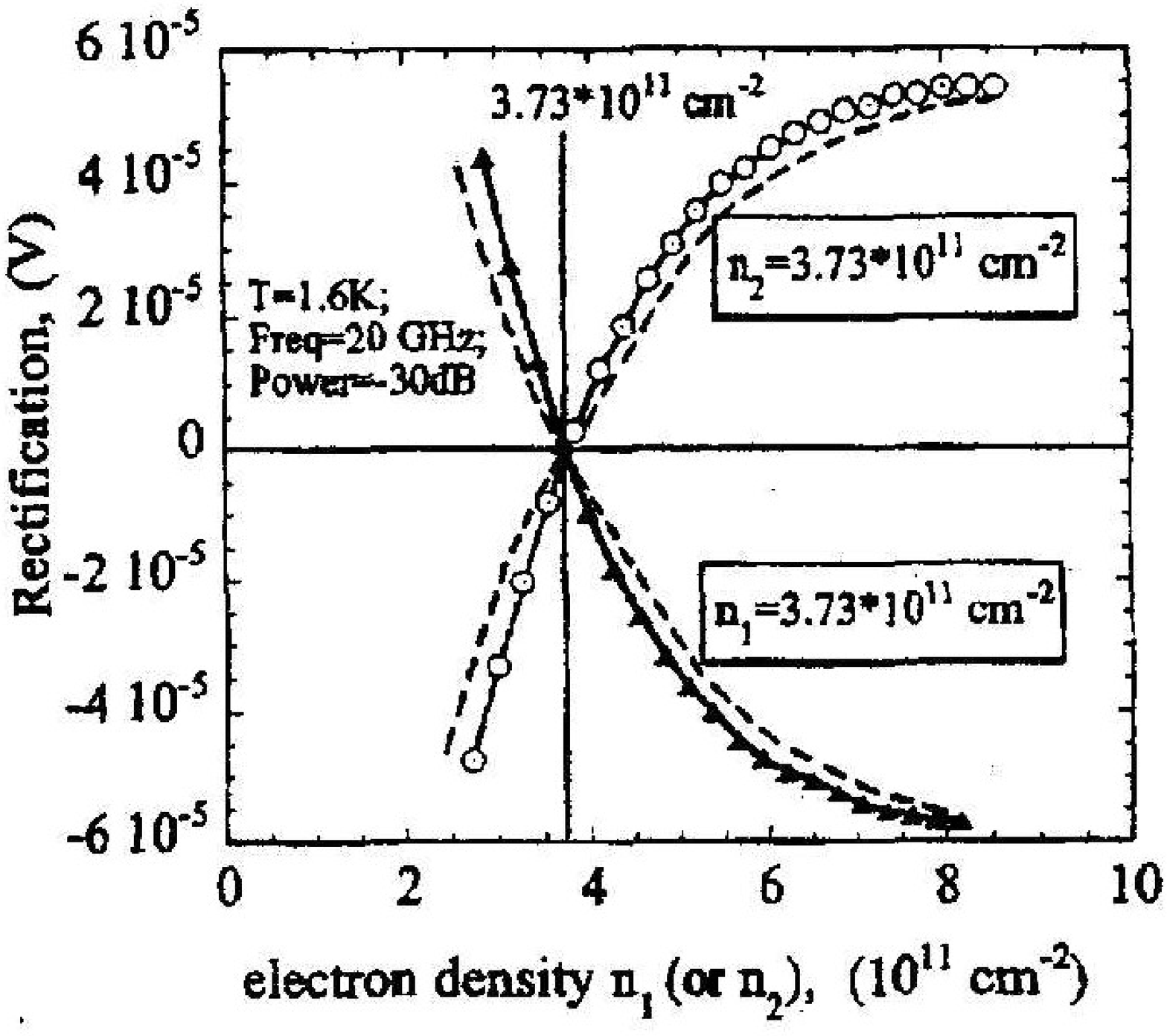}
\caption{
The rectified signal voltage $ V_{dc} $ is plotted versus
electron density. Solid lines -- experiment of Ref.[9]; dashed line --
theory for the parameters $ m \approx 0.2 m_0; \; \tau \approx 3 \times
10^{-12} s^{-1}.$
}
 \label{rateI}
\end{center}
\end{figure}

 In summary, obtained results based on the Boltzmann equation
exhibit the effect of the microwave rectification of the current in a 2D
electron gas. It is shown that the effect of rectification can originate
from the inhomogeneity of the electron density. The effect appears when an
asymmetric density gradient is created in the electron system by an
external static voltage applied to it. The dc current arises when the ac
electric field $ {\bf E} (t) $ has a nonzero component along the density
gradient. The results of theoretical analysis agree with the recently
reported experimental results [9].
 \vspace{2mm}
 
{\it Acknowledgments}.  I thank M.R. Sarachik for 
 enabling me to study the relevant experimental results and 
helpful discussions and G.M. Zimbovsky
for help with the manuscript.

 \vspace{2mm}

--------------------------------------------------------------

\vspace{2mm}

1. See e.g. F Julicher, A. Ajdari and J. Prost, Rev. Mod. Phys. {\bf 69},
1269 (1997).

2. See P. Reimann, Phys. Rep. {\bf 361}, 57 (2002), and referencies
therein.

3. F.G. Bass and Bulgakov, "Kinetic and Electrodynamic Phenomena in
Classical and Quantum Semiconductor Superlattices" (Nova Science
Publishing, Commack, NY, 1997).

4. V.I. Fal'ko, Sov. Phys. Solid State {\bf 31}, 561 (1989).

5. S. Flach, O. Yevtushenko, and Y. Zolotaryuk, Phys. Rev. Lett. {\bf 84},
2358 (2000).

6. J.L. Mateos,  Phys. Rev Lett. {\bf 84}, 258 (2000).

7. S. Flach, Y. Zolotaryuk, and A.A. Ovchinnikov Europhys.  Lett. {\bf
54}, 141 (2001).

8. S. Denisov, S. Flach, A.A. Ovchinnikov, O. Yevtushenko, and
Y. Zolotaryuk, Phys. Rev. E {\bf 66}, 041104 (2002).

9. I. Haxsa, S.A. Vitkalov, N.A. Zimbovskaya, M.P. Sarachik and T.M.  
Klapwijk, cond-mat/0110331.

10. K.W. Boer, ''Survey of Semiconductor Physics'', (Van Nostrand Reynold,
NY, 1990). 

11. I.O. Kulick, A.N. Omel'yanchuk, and R.I. Shekhter,
Sov. J. Low. Temp. Phys. {\bf 3}, 740 (1977).

12. A.P. van Gelder, Solid State Comm. {\bf 25}. 1097 (1978).

13. R.W. van der Hiijden, A.G.M. Jansen, J.H.M. Stoelinga, H.M. Swartjez,
and P. Wyder, Appl. Phys. Lett. {\bf 37}, 245 (1980).

14. We do not consider here in detail the relation of relaxation time
$\tau $ to the dissipation constant $ \gamma.$ In the present analysis we
only assume that the collisionless limit $\tau \to \infty $ corresponds to
the dissipationless limit $ \gamma = 0, $ and $\gamma $ takes on nonzero
values when $\tau $ is finite.

 \vspace{4mm}

\newpage

\end{document}